\documentclass[reprint,superscriptaddress,aps,prb]{revtex4-2}

\usepackage{graphicx}
\usepackage{hyperref}
\usepackage{changes}
\usepackage{makerobust}
\usepackage{xcolor}
\usepackage{physics}
\usepackage{amsmath}
\usepackage{amssymb}
\usepackage{mathtools}
\usepackage{enumitem}
\usepackage{orcidlink}

\newcommand{\bvec}[1]{\boldsymbol #1}
\begin{document}

\title{Magnetic phases of the anisotropic triangular Hubbard model from the ghost-Gutzwiller approximation in the rotating spin-frame}

\author{Azin~Kazemi-Moridani\,\orcidlink{0009-0008-9507-0541}}
\affiliation{Département de Physique et Institut Courtois, Université de Montréal, C.P. 6128, Succursale Centre-Ville,
Montréal, Québec, Canada H3C 3J7}
\affiliation{Center for Computational Quantum Physics, Flatiron Institute, New York, New York 10010, USA}

\author{Samuele~Giuli\,\orcidlink{0009-0004-7341-3655}}
\affiliation{Center for Computational Quantum Physics, Flatiron Institute, New York, New York 10010, USA}

\author{Tsung-Han~Lee\,\orcidlink{0000-0002-0571-9909}}
\affiliation{Department of Physics, National Chung Cheng University, Chiayi 62102, Taiwan}

\author{A.~-M.~S.~Tremblay\,\orcidlink{0000-0001-6932-8299}}
\affiliation{Département de physique \& Institut quantique \& RQMP Université de Sherbrooke, Sherbrooke, Québec, J1K 2R1 Canada}

\author{Michel~Côté\,\orcidlink{0000-0001-9046-9491}}
\affiliation{Département de Physique et Institut Courtois, Université de Montréal, C.P. 6128, Succursale Centre-Ville,
Montréal, Québec, Canada H3C 3J7}

\author{Nicola~Lanatà\,\orcidlink{0000-0003-0003-4908}}
\affiliation{School of Physics and Astronomy, Rochester Institute of Technology, Rochester, New York 14623, USA}
\affiliation{Center for Computational Quantum Physics, Flatiron Institute, New York, New York 10010, USA}

\author{Olivier~Gingras\,\orcidlink{0000-0003-3970-6273}}
\email{ogingras@flatironinstitute.org}
\affiliation{Center for Computational Quantum Physics, Flatiron Institute, New York, New York 10010, USA}
\affiliation{Université Paris-Saclay, CNRS, CEA, Institut de physique théorique, 91191, Gif-sur-Yvette, France}

\date{\today}

\begin{abstract}
    We investigate the magnetic phase diagram of the half-filled Hubbard model on the anisotropic triangular lattice using the Gutzwiller approximation (GA) and its ghost generalization (ghost-GA). By combining a rotating spin-frame formulation with high-resolution momentum grids, we determine magnetic ground states through direct total-energy minimization over the ordering wavevector. We benchmark standard GA and ghost-GA against dynamical mean-field theory (DMFT) and dual-fermion results.
    We show that GA already captures the qualitative structure of the phase diagram, but systematically overestimates the stability of magnetic order due to the absence of dynamical fluctuations. We find that introducing a small number of auxiliary “ghost” orbitals is sufficient to recover most dynamical effects and significantly improves quantitative agreement with DMFT. Exploring the full Brillouin zone, we obtain a phase diagram comprising paramagnetic and various magnetic phases. In contrast to ladder dual-fermion susceptibility-based predictions, we find that the one-dimensional antiferromagnetic phase is never stabilized, despite being the leading instability in certain regimes.
    Our results establish ghost-GA as an efficient and systematically improvable framework for studying magnetism in frustrated systems, capable of achieving near-DMFT accuracy at a fraction of the computational cost. They also highlight that standard GA performs qualitatively well for capturing the general phase diagram, enabling the investigation of incommensurate magnetic orders in more complex systems.
\end{abstract}

\maketitle

\section{Introduction}

The anisotropic triangular-lattice Hubbard model has long been used as a simple playground for studying the emergence of various ordered phases from the interplay between geometric frustration and strong local electronic repulsion. In this simple setting, antiferromagnetic interactions cannot be satisfied on all bonds at once, leading to a plethora of competing ordered states, including magnetic and spiral magnetic states, quantum spin liquids, and unconventional superconductivity.
Over the years, a variety of advanced numerical methods have been employed to investigate this model and the competing phases it can host, including dynamical mean-field theory (DMFT) and its cluster generalizations~\cite{PhysRevX.11.041013, PhysRevB.89.195108, PhysRevB.94.245133, PhysRevB.94.245145, PhysRevLett.133.136501, PhysRevX.12.021064, PhysRevLett.100.076402, PhysRevB.74.085117, PhysRevB.76.144513, PhysRevB.107.125159, PhysRevB.110.L121109, fournier2025frenkellinepseudogapanalogy, Sahebsara_Senechal_2008, Kyung_Tremblay_2006}, ladder dual-fermion approximation~\cite{PhysRevB.107.075106, PhysRevB.91.245125}, quantum Monte Carlo (QMC)~\cite{PhysRevB.80.064419, Watanabe2006,PhysRevB.77.214505,  Meng2010, PhysRevB.87.035143, PhysRevB.91.245125, PhysRevX.11.041013}, tensor networks~\cite{PhysRevB.103.235132, PhysRevX.10.021042, PhysRevX.11.041013, PhysRevResearch.4.043048}, slave-particle approaches~\cite{Arrigoni1991,feiguin1997SB,Capone2001} and the renormalization group~\cite{Tsai_2001}, but also analogue, platforms such as silicon platforms~\cite{PhysRevLett.119.266802} and cold atoms quantum simulators~\cite{PRXQuantum.2.020344, Xu2023Frustration, Lebrat2024}.

Not only does this model capture a fascinating competition between multiple states, but it is also believed to include the central microscopic interactions underlying the rich phase diagrams of a wide range of frustrated materials. For example, it is often employed to explain and investigate the behavior of realistic systems including organic charge-transfer salts such as the $\kappa$-(BEDT-TTF)$_2$X family~\cite{PhysRevB.74.085117, Tsai_2001, PhysRevLett.100.076402, PhysRevB.91.245125, PhysRevB.80.064419, Geiser1991, Geiser1991kappaBEDT, Ivanov1997Electronic, Urayama1988Crystal, Watanabe2006}, low-energy models for moiré materials~\cite{Kennes2021, Wang2020, Tang2020, PhysRevResearch.4.043048, PhysRevLett.121.026402, PhysRevLett.122.086402, PhysRevX.12.021064, PhysRevLett.121.026402, qdqb-w7zl}, 1T-TaS$_{2-x}$Se$_x$~\cite{PhysRevLett.112.206402, PhysRevLett.97.067402} although a multi-orbital description might be necessary~\cite{PhysRevX.7.041054, Dalal2025, Pizarro2020, Crippa2024}, along with Li$_x$NbO$_2$~\cite{PhysRevB.105.104504, PhysRevB.76.144513} and many other systems.

Despite decades of work, obtaining a definitive phase diagram remains a challenge, in particular regarding quantum spin liquids. 
However, for the purely magnetic phase diagram, a certain consensus between the methods is starting to arise.
While the precise phase boundaries vary from method to method, the main competing states are as expected: a paramagnet at small interaction strength and various magnetic orders at larger strength, from a Néel-like order at weak frustration, a spiral order at isotropic frustration, decoupled antiferromagnetic chains at large frustration, and more complex non-collinear arrangements in between them.

Among the numerous numerical methods employed to investigate the anisotropic triangular Hubbard model, DMFT is a natural approach to study its magnetic phase diagram because it treats the local Hubbard interaction non-perturbatively while remaining unbiased with respect to magnetic ordering. Magnetism in this model is driven by the formation and ordering of local moments, which DMFT captures exactly in the infinite-dimension limit~\cite{PhysRevLett.62.324, PhysRevB.45.6479, RevModPhys.68.13, RevModPhys.78.865, https://doi.org/10.1002/andp.201100250}. The effects of hopping anisotropy enter solely via the embedding of the local electronic correlations back onto the lattice, allowing changes in bandwidth and effective dimensionality to be incorporated without additional approximations. 
However, there are two important drawbacks of  DMFT for investigating the magnetic phase diagram of the triangular Hubbard model.

The first one is the numerical cost of DMFT.
This motivates the use of the ghost-Gutzwiller approximation (ghost-GA), which has been shown to reproduce DMFT results with good accuracy while incurring significantly reduced numerical cost~\cite{PhysRevB.107.L121104, MejutoZaera2023_efficient, PhysRevB.108.245147, makaresz2026acceleratingdynamicalmeanfieldtheory}, particularly when combined with machine learning solvers~\cite{PhysRevResearch.6.013242, zhou2025neuralquantumstatesimpuritysolverquantum, giuli2025linearfoundationmodelquantum}.
The ghost-GA is a variational approach that extends the conventional Gutzwiller wavefunction framework by introducing auxiliary (“ghost”) degrees of freedom that capture dynamical correlation effects beyond static renormalization factors. This extension enables a more accurate description of correlated metallic and magnetic phases while retaining a computational complexity comparable to that of mean-field methods~\cite{PhysRevB.85.035133, PhysRevB.96.195126, PhysRevB.104.L081103,Giuli2025alm,bellomia2025}. As a result, the ghost-GA provides an efficient and reliable alternative for exploring magnetic phase diagrams in strongly correlated lattice models. Moreover, while including zero ghosts yields the standard Gutzwiller approximation (GA)~\cite{PhysRevLett.10.159, PhysRev.134.A923, PhysRev.137.A1726}, a systematic improvement of the method is possible by increasing the number of ghosts, and in the limit of infinite ghosts, the DMFT solution is recovered~\cite{Giuli2026unifying}.

This second drawback is not specific to DMFT, but is also shared by local embedding approaches such as ghost-GA. To capture a general magnetically ordered state, one must be able to stabilize a finite staggered magnetization, which locally on site $i$ is expressed as $\textbf{M}_i = \langle \sum_{\sigma \sigma'} \hat{c}^\dagger_{i\sigma} \frac{\bvec{\sigma}_{\sigma \sigma'}}{2} \hat{c}_{i\sigma'} \rangle$, where $\hat{c}_{i\sigma}$ ($\hat{c}^\dagger_{i\sigma}$) annihilates (creates) an electron with spin $\sigma$ at site $i$ and $\bvec{\sigma}$ is a vector of Pauli matrices. The spatial modulation of $\textbf{M}_i$ is encoded in its Fourier transform $\textbf{M}_{\textbf{q}} \equiv \int d\bvec{r_i} \ e^{i\textbf{q} \cdot \textbf{r}_i} \ \textbf{M}_i$, with $\textbf{r}_i$ the position of site $i$ in real space. A finite wavevector $\mathbf{q}$ describes how the spin orientation rotates from site to site. Such modulated states break both the $SU(2)$ spin symmetry and the translational symmetry of the original lattice. For local embedding approaches, the latter is the main difficulty. As shown in previous works, this issue can be overcome by performing a local $SU(2)$ rotation that transfers the $\mathbf{q}$-dependence from the order parameter to the hopping amplitudes~\cite{PhysRevLett.61.467, Arrigoni1991, PhysRevLett.80.2393, Cote_Tremblay_1995, PhysRevB.60.5224, PhysRevB.94.245145, Riegler_2020,Klett_2020}.
It is also known as the generalized Bloch theorem~\cite{Sandratskii_1991}.
We refer to this formulation as the rotating spin-frame. The resulting Hamiltonian is translationally invariant for each chosen $\mathbf{q}$, enabling the evaluation of observables--including total energies--within impurity-based methods.

In this work, we exploit the rotating spin-frame formulation to compute the total energy landscape $E_\mathbf{q}$ with high-momentum \textbf{q}-grids resolution for both GA and ghost-GA, enabling us to track the continuous evolution of incommensurate magnetic orders across variations of the model parameters. We focus on the Hubbard model on the anisotropic triangular lattice at half-filling, with the Coulomb repulsion $U/t$ (where $t$ is the nearest-neighbor hopping amplitude) and the geometric frustration $t'/t$ as parameters. We report the $(U/t,t'/t)$ phase diagrams and compare with results from DMFT~\cite{PhysRevB.94.245145} and ladder dual-fermion (DF)~\cite{PhysRevB.107.075106}. The manuscript is organized as follows: The model, classification of magnetic states, and methodological details are presented in Sec.~\ref{sec:model_method}.
In Sec.~\ref{sec:results}, we present and interpret our results, which show that:
\begin{enumerate}[label=(\roman*)]
    \item GA reproduces the qualitative structure of the magnetic phase diagrams obtained in DMFT and ladder DF, though it quantitatively overestimates the stability of ordered phases since it lacks dynamical and local fluctuations.
    \item Introducing ghost orbitals systematically and significantly improves the agreement with DMFT by partially restoring the dynamical fluctuations.
    \item Neither GA nor ghost-GA stabilizes the one-dimensional antiferromagnetic state inferred from DF susceptibilities~\cite{PhysRevB.107.075106}. By investigating the magnetic susceptibility, we find that, although this state is the leading magnetic instability, it is not stabilized in ghost-GA. This indicates that, in this system, a susceptibility maximum alone does not guarantee that the corresponding order is stabilized as the ground state.
\end{enumerate}
We conclude in Sec.~\ref{sec:concl} by affirming that GA provides a simple, robust, and computationally advantageous tool to efficiently investigate magnetically ordered ground states in correlated systems, and that the addition of ghost variational degrees of freedom offers a clear path toward systematically improving accuracy. 
Their efficiency also makes them promising for future applications to more complex systems, including multi-orbital models such as Sr$_2$FeO$_4$~\cite{PhysRevB.109.165146}, cluster extensions, and superconducting states, where dense momentum scans would otherwise be computationally prohibitive.

\section{Model \& Method}
\label{sec:model_method}

\begin{figure}
    \centering
    \includegraphics[width=1\linewidth]{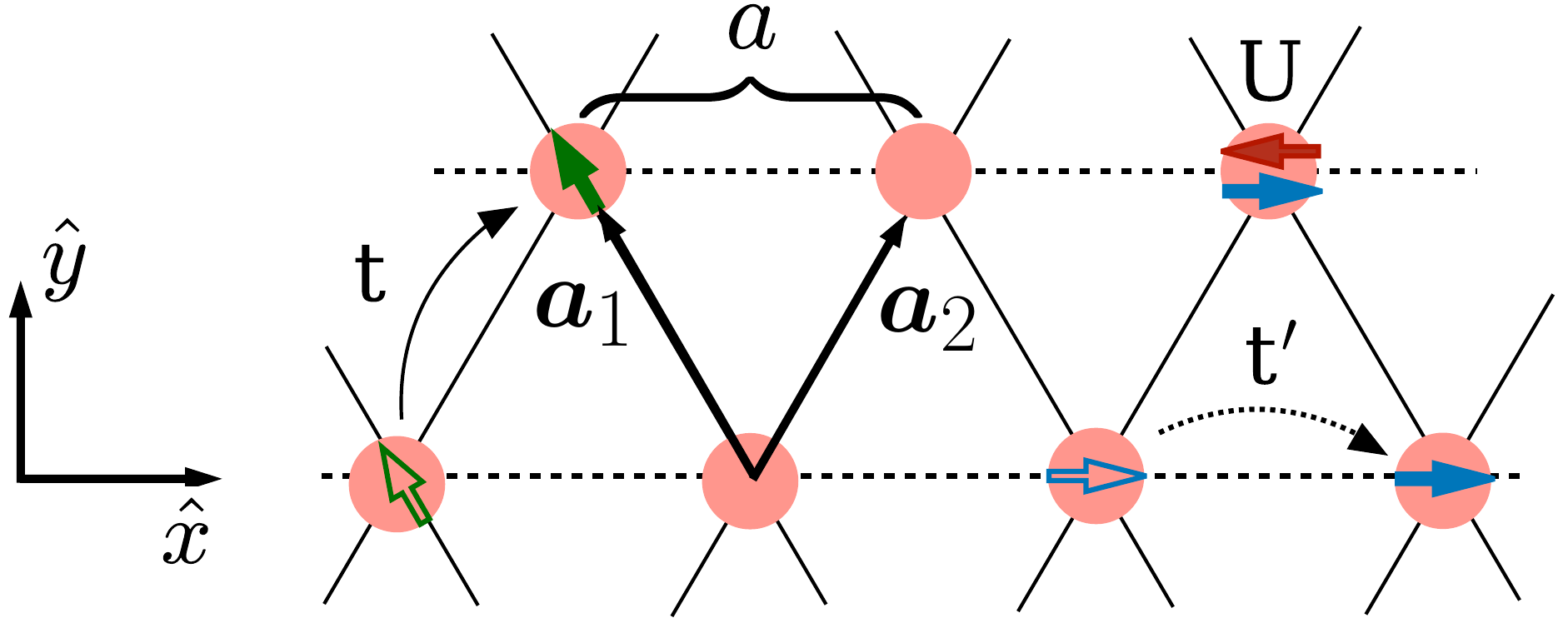}
    \caption{
    Hubbard model and the geometry of the anisotropic triangular lattice. The model contains the nearest-neighbor hoppings $t$ along $\pm\bvec{a}_1$ and $\pm\bvec{a}_2$, and $t^\prime$ along $\pm(\bvec{a}_1-\bvec{a}_2)$, along with the double occupation cost $U$. For $t^\prime=t$, the model is isotropic and $C_6$ symmetric, and anisotropic for $t^\prime\neq t$.}
    \label{fig:model}
\end{figure}

\begin{figure*}[!htbp]
    \includegraphics[width=0.85\textwidth]{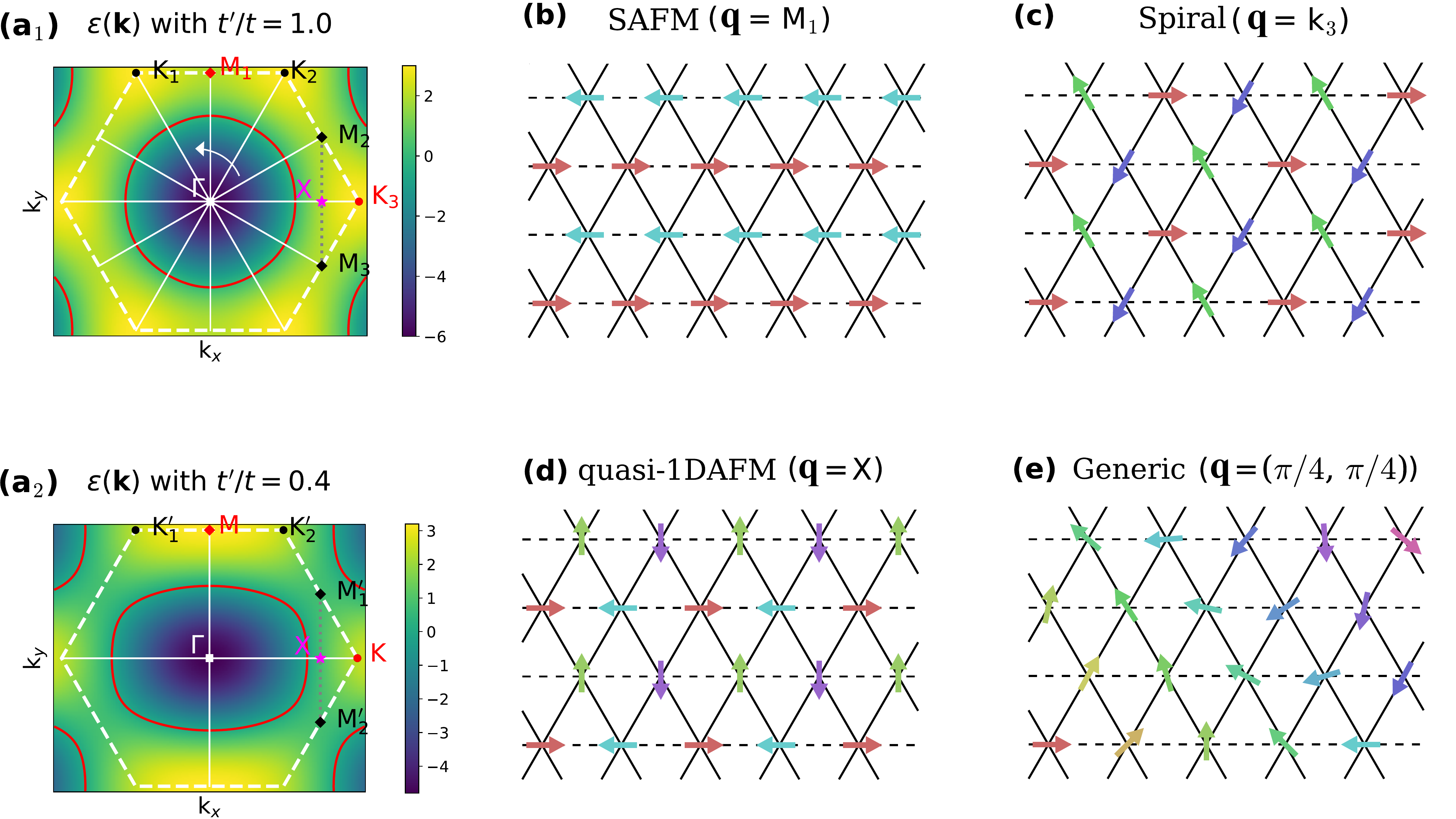}
    \caption{
    (a$_1$, a$_2$)~Non-interacting dispersions $\varepsilon$($\bvec{k}$) of the triangular lattice for the isotropic ($t^\prime/t = 1$) and the anisotropic (here $t^\prime/t = 0.4$) lattice, respectively. The dashed hexagon, full straight lines, and arrow indicate the Brillouin zone, mirror symmetries, and $C_6$ symmetry, respectively. Solid red lines indicate the Fermi surface. The dotted grey line highlights the momentum characteristic of decoupled one-dimensional chains. In the isotropic case, the $M$ and $K$ points are symmetry-equivalent. In the anisotropic case, the $M'$ and $K'$ points are symmetry-equivalent, but inequivalent to the $M$ and $K$ points.
    (b-e)~Real space representation of the different classes of magnetic phases.
    (b)~Square antiferromagnet (SAFM), characterized by a collinear Néel order with ordering vector at the $M$ point.
    (c)~Spiral order, with ordering vector at the $K$ point, corresponding to the $120^\circ$ coplanar spin structure realized in the isotropic triangular lattice limit.
    (d)~ Quasi one-dimensional antiferromagnetic order (quasi-1DAFM), stabilized in the large $t'/t$ limit and characterized by antiferromagnetic correlations along a single lattice direction. 
    (e)~Generic spin-density wave, where the ordering vector lies away from high-symmetry points, leading to a continuously varying spin angle.
    }
    \label{fig:magnetic_orderings}
\end{figure*}

We study the magnetic ground-state of the Hubbard model on the anisotropic triangular lattice, which is described by the Hamiltonian depicted in Fig.~\ref{fig:model}. It has the form $\hat{\mathcal{H}} = \hat{\mathcal{H}}_0 + \hat{\mathcal{H}}_U$ where $\hat{\mathcal{H}}_0$ is the non-interacting part, defining the kinetics of the electrons, and $\hat{\mathcal{H}}_U$ is the Hubbard interaction, which captures the on-site Coulomb repulsion. They are respectively given by
\begin{equation}
    \label{eq:Hubb_model}
    \hat{\mathcal{H}}_0 = \sum_{ij\sigma} t_{ij} \hat{c}^\dagger_{i\sigma} \hat{c}_{j\sigma} - \mu \sum_{i\sigma} \hat{n}_{i\sigma},
    \ \
    \hat{\mathcal{H}}_U = U \sum_{i} \hat{n}_{i\uparrow} \hat{n}_{i\downarrow},
\end{equation}
where $\hat{c}^\dagger_{i\sigma}$ ($\hat{c}_{i\sigma}$) creates (annihilates) an electron at site $i$ with spin $\sigma$, $\hat{n}_{i\sigma} \equiv \hat{c}^\dagger_{i\sigma} \hat{c}_{i\sigma}$ is the associated density operator, $t_{ij}$ is the hopping amplitude from site $j$ to site $i$, $U$ is the on-site Coulomb repulsion and $\mu$ is the chemical potential. Half-filling is enforced by converging the paramagnetic solution on the electronic density.

To study the anisotropic triangular lattice also represented in Fig.~\ref{fig:model}, we only consider nearest-neighbor hoppings, where the cost for an electron to hop between two sites connected by $\pm \bvec{a}_1$ or $\pm \bvec{a}_2$ is $t$, and the cost for neighbors connected by $\pm \bvec{a}_1 \mp \bvec{a}_2$ is $t'$. The resulting dispersion is shown in Fig.~\ref{fig:magnetic_orderings} for two different values of $t'/t$. 
In Fig.~\ref{fig:magnetic_orderings}(a$_1$), for $t' = t$, the lattice is $C_6$ symmetric and said to be isotropic. In this case, the $M_i$ points are related by symmetry, and so are the $K_i$ points. The full white lines represent mirror symmetries. In Fig.~\ref{fig:magnetic_orderings}(a$_2$), for $t' \neq t$, the lattice is said to be anisotropic. Consequently, the three $M$ ($K$) points are split into one $M$ ($K$) and two $M'$ ($K'$) points. There are much fewer mirror symmetries. Throughout this manuscript, we fix $t$ and investigate the phase diagram by varying $t'$ and $U$. Moreover, note that this geometry is equivalent to considering the square lattice, with an additional diagonal term, as was done in Ref.~\citenum{PhysRevB.94.245145}. We show in App.~\ref{app:triangle-to-square} how to map one lattice to the other.

In particular, in this manuscript, we are interested in the magnetic phase diagram of this model in the $(U/t,t'/t)$ parameter space. 
This section is organized as follows.
In Sec.~\ref{sec:model_method-classification}, we introduce the staggered magnetization, the magnetic order parameter, and discuss how its $\textbf{q}$-dependence is used to classify the different states of the phase diagram.
In Sec.~\ref{sec:model_method-frame}, we explain the rotating spin-frame formulation, which allows us to capture the $\textbf{q}$-phase modulation of the local magnetization within a single cell. The drawback of this formulation is that we need to perform a calculation for every $\textbf{q}$-point independently. We therefore first converge a paramagnetic calculation at half-filling, and from that solution explore a dense \textbf{q}-grid, to resolve various incommensurate magnetic orders.
In Sec.~\ref{sec:model_method-gGA}, we present the ghost-GA method (for which standard GA is a special case), which is employed throughout this work.

\subsection{Classification of the magnetic phases}
\label{sec:model_method-classification}

Although the Hubbard model in Eq.~(\ref{eq:Hubb_model}) is invariant under global $SU(2)$ spin rotations, magnetically ordered states spontaneously break this symmetry and, in a symmetry-broken description, are characterized by a local magnetization with arbitrary orientation in spin space.
In this work, following common practice for this system, we restrict our analysis to coplanar magnetic configurations, i.e., states in which all local magnetization vectors lie within a common plane. Owing to the underlying $SU(2)$ symmetry, this plane can be chosen without loss of generality to be the $xy$-plane. Extension to fully three-dimensional spin configurations is straightforward.

With this assumption, geometric frustration and interaction dictate how the direction of the local magnetization rotates in the plane as a function of spatial position. This rotation is encoded by a wavevector $\mathbf{q}$ and the associated Fourier component is given by
\begin{equation}\label{eq:order_parameter}
    M_{\textbf{q}} = \frac{1}{N} \sum_i \Big\langle \hat{S}_{ix} \cos (\textbf{q} \cdot \textbf{r}_i) + \hat{S}_{iy} \sin (\textbf{q} \cdot \textbf{r}_i ) \Big \rangle,
\end{equation}
where $\mathbf{r}_i$ denotes the position of site $i$ in real space, $N$ is the number of lattice sites, and
\begin{align}
    \hat{S}_{ix} &\equiv \frac{1}{2}\bigl(
    \hat{c}^\dagger_{i\uparrow}\hat{c}_{i\downarrow} 
    + 
    \hat{c}^\dagger_{i\downarrow}\hat{c}_{i\uparrow}
    \bigr), \quad
    \hat{S}_{iy} \equiv -\frac{i}{2}\bigl(
    \hat{c}^\dagger_{i\uparrow}\hat{c}_{i\downarrow} 
    - 
    \hat{c}^\dagger_{i\downarrow}\hat{c}_{i\uparrow}
    \bigr)
\end{align}
are in-plane spin operators, expressed in terms of spin-flip operators.

The nature of the magnetic order is determined by the ordering wavevector $\mathbf{q}$ for which the $M_{\mathbf{q}}$ is finite, and the total energy $E_\textbf{q}$ is minimized. 
Different locations of this wavevector \textbf{q} in the Brillouin zone correspond to distinct magnetic phases, set by the interplay between electronic correlations and geometric frustration of the lattice. Throughout this work, we use five labels to categorize the different phases that we find. Some representative examples of real-space spin structures are presented in Fig.~\ref{fig:magnetic_orderings}. With the associated labels, they are:
\begin{enumerate}[label=(\alph*)]
    \item Paramagnet (PM): The state with the lowest total energy has no finite magnetization on the lattice. As a consequence, the energy $E_\textbf{q}$ is independent of \textbf{q}.
    \item Square antiferromagnet (SAFM): In the $t'/t \to 0$ limit, the triangular lattice is equivalent to the square lattice (see App.~\ref{app:triangle-to-square}). At half-filling, the magnetization is in a checkerboard pattern. This ordering is associated to $\mathbf{q}$ at the $M$ point in Fig.~\ref{fig:magnetic_orderings}(a$_2$).
    \item Spiral order: At $t=t^\prime$, geometric frustration is maximal and the local moments prefer a $120^\circ$ angle with each other, leading to a spiral state. This ordering corresponds to $\mathbf{q}$ at the $K$ points in Fig.~\ref{fig:magnetic_orderings}(a$_1$).
    \item One-dimensional antiferromagnet (1DAFM): The strongly anisotropic regime $t' \gg t$ describes the lattice as a set of weakly coupled antiferromagnetic chains running along the $\mathbf{a}_1-\mathbf{a}_2$ direction. In the limit $t \to 0$ at finite $U$ and $t'$, the total energy becomes independent of $\mathbf{q}$ along the line connecting $M'_1$ and $M_2'$ in Fig.~\ref{fig:magnetic_orderings}(a$_2$), reflecting an extensive degeneracy associated with decoupled one-dimensional chains.
    \item Generic spin-density wave: Away from these specific limits ($t'/t \to 0, 1, \infty$), a magnetic state can be found at a generic $\mathbf{q}$ away from high-symmetry points.
\end{enumerate}
Resolving these magnetic tendencies therefore requires access to the full momentum dependence of the order parameter $M_\textbf{q}$.
In conventional single-site quantum embedding theories such as DMFT and ghost-GA, such momentum resolution is not directly available because translational symmetry is enforced when embedding the impurity self-energy back onto the lattice. 

One route to overcome this limitation is to analyze magnetic susceptibilities from the high-temperature normal state and extract the dominant ordering wavevector from their \textbf{q}-dependence via the Bethe--Salpeter equation~\cite{PhysRevB.107.075106}. 
An alternative approach, adopted in this work and in Ref.~\citenum{PhysRevB.94.245145}, is to perform a local $SU(2)$ gauge transformation that absorbs the spatial dependence of the magnetization into the non-interacting Hamiltonian. 
This spin-rotating frame allows us to capture the continuous evolution of the magnetic states with arbitrary ordering wavevector $\mathbf{q}$ and to determine the \textbf{q} that minimizes the total energy. This approach is detailed in the following subsection.

\subsection{Rotating spin-frame formulation}
    \label{sec:model_method-frame}    

    The staggered magnetization $M_{\textbf{q}}$ defined in Eq.~(\ref{eq:order_parameter}) breaks the translational invariance of the lattice.
    As a result, local quantum embedding methods such as DMFT or ghost-GA cannot capture this ordering unless one allows for site-dependent (inequivalent) embeddings, which would require solving and converging an extensive number of impurity problems. 
    A more convenient approach is to use a suitable unitary transformation, recasting the problem into a form where both the local Hamiltonian and the magnetic ordering remain translationally invariant.
    Indeed, the Hamiltonian given by Eq.~(\ref{eq:Hubb_model}) can instead be mapped to one where the translational modulation of the order parameter is absorbed by its non-interacting part.
    
    To see this, consider applying the following spin- and position-dependent gauge transformation to the creation and annihilation operators:
    \begin{equation}
        \hat{c}^\dagger_{i\sigma} = e^{i\sigma(\textbf{q} \cdot \textbf{r}_i)}\hat{c}^\dagger_{\textbf{q}i\sigma} \quad \text{and} \quad \hat{c}_{i\sigma} = e^{-i\sigma(\textbf{q} \cdot \textbf{r}_i)}\hat{c}_{\textbf{q}i\sigma},
    \end{equation}
    where $\sigma = \uparrow$ ($\downarrow$) is treated as $+\frac{1}{2}$ ($-\frac{1}{2}$).
    In this rotated basis that depends on \textbf{q}, the interaction part of the Hamiltonian is untouched as
    \begin{equation}
        \hat{\mathcal{H}}_U \rightarrow \hat{\mathcal{H}}_{\textbf{q}U} = U \sum_{i} \hat{n}_{\textbf{q}i\uparrow} \hat{n}_{\textbf{q}i\downarrow},    
    \end{equation}
    where $\hat{n}_{\textbf{q}i\sigma} \equiv \hat{c}^\dagger_{\textbf{q}i\sigma}\hat{c}_{\textbf{q}i\sigma}$,
    but the tight-binding part becomes
    \begin{equation}
        \label{eq:ham0_q}
        \hat{\mathcal{H}}_{\textbf{q}0} = \sum_{ij\sigma} t_{ij} e^{i\sigma \textbf{q} \cdot (\textbf{r}_i - \textbf{r}_j)}\hat{c}^\dagger_{\textbf{q}i\sigma} \hat{c}_{\textbf{q}j\sigma} - \mu \sum_{i\sigma} \hat{n}_{\textbf{q}i\sigma}.
    \end{equation}
    In this basis, the magnetization Eq.~(\ref{eq:order_parameter}) becomes
    \begin{equation}
        M_\textbf{q} = \frac{1}{N} \sum_i \Big \langle \hat{S}_{\textbf{q}ix} \Big \rangle = \frac{1}{2N} \sum_i \Big \langle \hat{c}^\dagger_{\textbf{q}i\uparrow}\hat{c}_{\textbf{q}i\downarrow} + \hat{c}^\dagger_{\textbf{q}i\downarrow}\hat{c}_{\textbf{q}i\uparrow} \Big \rangle,
    \end{equation}
    which is now translationally invariant and can therefore be captured within single-site, translationally invariant DMFT and ghost-GA, given that spin-off-diagonal elements are allowed.

\subsection{Ghost-Gutzwiller approximation}
\label{sec:model_method-gGA}

The standard Gutzwiller approximation (GA) is a variational framework based on an auxiliary non-interacting Slater determinant defined in a Hilbert space of the same size as the physical problem~\cite{PhysRevLett.10.159, PhysRev.134.A923, PhysRev.137.A1726, PhysRevB.85.035133}. Electronic correlations are incorporated in a local projection operator acting on this wavefunction, such that all variational degrees of freedom are ultimately encoded in the one-body density matrix. While this construction successfully captures low-energy quasiparticle renormalization, it is intrinsically limited to describing incoherent spectral weight, excited states, and the multiplet structure characteristic of Mott and Hund physics.

The ghost-Gutzwiller approximation (ghost-GA) overcomes these limitations by enlarging the auxiliary Hilbert space~\cite{Lanata2017, pavarini2023orbital}. Specifically, it introduces $B-1$ additional auxiliary (“ghost”) orbitals per physical orbital, such that the total auxiliary space contains $B$ orbitals per physical degree of freedom, with $B=1$ recovering standard GA. This enlarged variational space allows for a much richer description of correlations, including incoherent spectral features and dynamical effects. In particular, the resulting variational self-energy acquires a finite number of poles, whose number is controlled by $B$, enabling a systematic improvement of the description of electronic excitations.

Recently, it has been shown that in the limit $B \rightarrow \infty$, ghost-GA becomes equivalent to DMFT~\cite{Giuli2026unifying}. Importantly, this equivalence is achieved within a variational framework that depends only on the one-body density matrix, in contrast to DMFT, which requires the full frequency-dependent Green’s function. This makes ghost-GA computationally much more efficient while retaining a controlled connection to DMFT.
In practice, only a small number of ghost orbitals per physical orbital is sufficient to obtain results in close quantitative agreement with DMFT~\cite{Lanata2017, PhysRevB.99.115129, PhysRevB.105.045111, PhysRevB.107.L121104, makaresz2026acceleratingdynamicalmeanfieldtheory}.

\begin{figure*}
    \includegraphics[width=\textwidth]{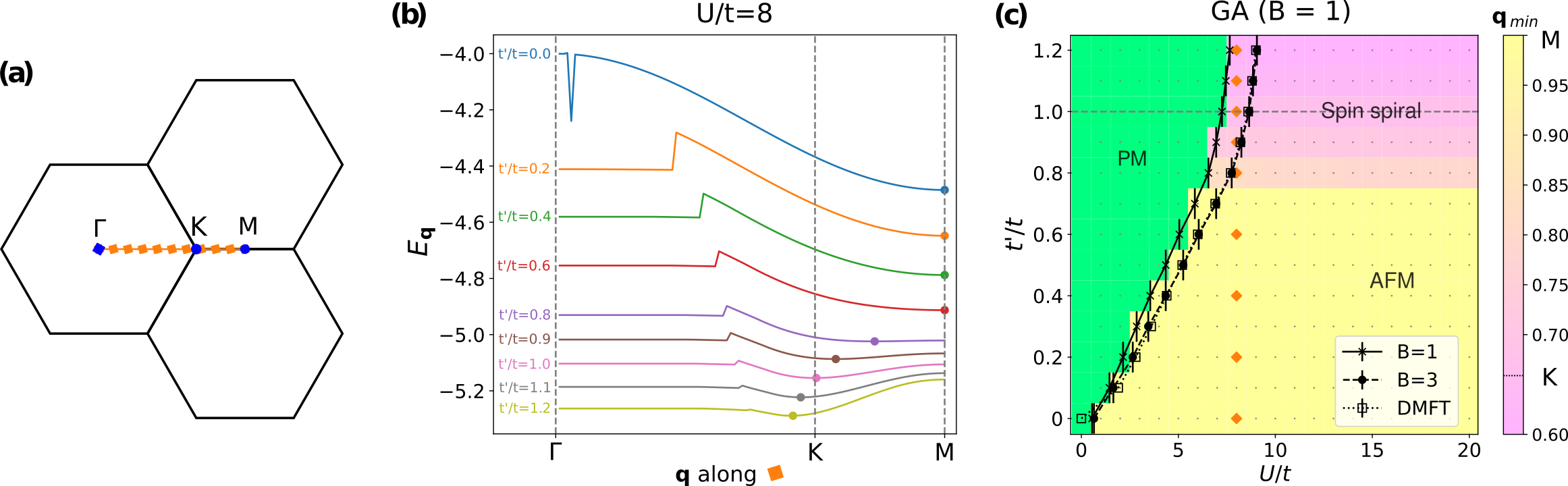}
    \caption{
        Results for magnetic phases with a wavevector along the $\Gamma-M$ high-symmetry path.
        (a)~Visual representation of the high-symmetry path. It goes from the zone center at $\Gamma$ to the middle of a neighboring zone edge at $M$, going through a corner of the Brillouin zone at $K$. This path allows us to capture both the AFM state (at $M$), the spiral state (at $K$), the quasi-one-dimensional 1DAFM state (at $X$ in the middle), and other generic states in between.
        (b)~Total energy $E_{\textbf{q}}$ in the grand canonical ensemble along the path $\Gamma-M$ at fixed $U/t=8$ for several $t^\prime/t$ obtained with GA ($B=1$). All states with finite $t'$ display a plateau at small momentum, where the system falls into the metallic solution. The sharp discontinuity signals that we were able to stabilize the magnetic state. Orange diamonds depict these points on the phase diagram in (c). The minimum in $E_\textbf{q}$ shifts as geometric frustration increases.
        (c)~Resulting GA phase diagram. The green color represents the paramagnetic (PM) phase, while the color gradient highlights that we can capture the continuous evolution in \textbf{q}, from AFM in yellow to quasi-1DAFM in pink, as we increase $t^\prime/t$. We also performed calculations on a finer ($\Delta U =0.1$) grid around the critical $U$ to obtain the $B=1$ and $B=3$ phase boundaries, depicted in full and dashed black lines, respectively. We also extracted the DMFT phase boundary from Ref.~\citenum{PhysRevB.94.245145} to compare with ghost-GA. 
    }
    \label{fig:1D_results}
\end{figure*}

In this work, we do not rederive the full ghost-GA formalism, which is presented in Refs.~\cite{Lanata2017, pavarini2023orbital, Giuli2026unifying}.
The calculations were performed using an implementation developed within the TRIQS library~\cite{PARCOLLET2015398}, which will be released separately.
The non-interacting Hamiltonian $\hat{\mathcal{H}}_0$ is Fourier transformed to a $413 \times 413$ \textbf{k}-grid, half-filling in the paramagnetic solution is converged to a precision of $10^{-5}$ and the variational parameters to a precision of $10^{-6}$, with an effective inverse temperature of $\beta=500$. The \textbf{q}-path used in the one-dimensional phase diagram Fig.~\ref{fig:1D_results} contains $100$ points from $\Gamma$ to $M$ along $K$, the \textbf{q}-grid for the two-dimensional phase diagram Fig.~\ref{fig:2D_vs_Gull} contains $57 \times 49$ points, and the susceptibilities of Fig.~\ref{fig:2d-susc} are calculated on a $29 \times 25$ \textbf{q}-grid. Each calculation runs for a few minutes on a single central processing unit (CPU). Converging the chemical potential is the limiting factor in the paramagnetic state and converging each $\textbf{q}$-point takes on the order of a few minutes.

\section{Results \& Discussion}
\label{sec:results}

In this section, we present the magnetic phase diagrams of the anisotropic triangular Hubbard model obtained with ghost-GA for both $B=1$ (standard GA) and $B=3$.
For each pair of interaction strength $U/t$ and hopping $t^\prime/t$, we determine the magnetic ground state by computing the total energy $E_\textbf{q}$ in the rotating spin-frame for a path or a grid of ordering wavevectors $\mathbf{q}$.  
In this formulation, the $\mathbf{q}$-dependence enters the non-interacting Hamiltonian, and the GA or ghost-GA variational parameters are optimized for each $\mathbf{q}$.  
The global minimum of $E_\textbf{q}$ identifies the preferred magnetic ordering vector given that the magnetization is finite.

In Sec.~\ref{sec:results_1d}, we explore the phase diagram obtained when investigating magnetic instabilities along the $\Gamma-K-M$ high-symmetry path, as was done in Ref.~\citenum{PhysRevB.94.245145} with DMFT.
Our results suggest that GA already provides a qualitatively correct description of the phases, but their boundaries differ from DMFT. Employing ghost-GA instead, the boundaries become hardly distinguishable from DMFT with only $B=3$.

In Sec.~\ref{sec:results_2d}, we present calculations performed on the full two-dimensional Brillouin zone and compare our phase diagram with that obtained using ladder DF reported in Ref.~\citenum{PhysRevB.107.075106}. 
We find a substantially different phase diagram. This difference can be traced to two methodological differences. First, the analysis reported in Ref.~\citenum{PhysRevB.107.075106} is based on the magnetic susceptibility rather than on the direct stabilization of the ordered phase. Second, it includes additional non-local fluctuations that are absent in our calculations.
In particular, we never stabilize, at small $U/t$, the 1DAFM phase that they report, and our calculations of magnetic susceptibilities suggest we never will.

\subsection{One-dimensional phase diagram}
\label{sec:results_1d}

We first investigate incommensurate magnetism with a momentum \textbf{q} along the high-symmetry path depicted in Fig.~\ref{fig:1D_results}(a). This is analogous to the path investigated in Ref.~\citenum{PhysRevB.94.245145} with DMFT at zero temperature. Although a spin-density wave with a generic \textbf{q}-modulation outside of this path is possible, as discussed in Sec.~\ref{sec:results_2d}, it turns out that this high-symmetry path captures the essential phase diagram of this system. 

For each point in the $(U/t,t'/t)$-space, we first perform a ghost-GA ($B=1$ and $B=3$) paramagnetic calculation in order to converge the chemical potential $\mu$ to enforce half-filling, and the ghost-GA variational parameter matrices that capture local correlations. From those initial conditions, we scan the high-symmetry path of Fig.~\ref{fig:1D_results}(a) by performing the rotating spin-frame transformation detailed in Sec.~\ref{sec:model_method-frame}, for each \textbf{q} along this line. Figure~\ref{fig:1D_results}(b) shows examples of the results for the total energy $E_\textbf{q}$ curves for $U/t=8$ and for various $t'/t$ values. We observe that our results reproduce the qualitative behavior found in DMFT~\cite{PhysRevB.94.245145}: in the square lattice regime ($t'/t=0$), the system prefers an AFM state (minimum at the $M$ point) with alternating spins along the dominant hopping directions. As frustration $t'/t$ increases, the minimum initially remains at the $M$ point, until $t'/t$ becomes substantial and the minimum moves towards the $K$-point. When $t'/t$ reaches $1$, the system is six-fold symmetric and the \textbf{q}-point with lowest energy is at $K$, referred to as the spiral state. As $t'$ becomes larger than $t$, the minimum continues to move towards the $\Gamma$ point, but actually saturates halfway between $\Gamma$ and $M$ at the $X$ point, where the magnetic state resembles a 1DAFM of completely decoupled chains.

Performing these scans for $U/t \in [0, 20]$ and extracting the $\textbf{q}$-point with lowest $E_\textbf{q}$ yields the phase diagram presented in Fig.~\ref{fig:1D_results}(c). The colors correspond to the states obtained on a coarse-grain grid of $(U/t, t'/t)$ with GA. The situation where the $\text{q}$-point with the lowest total energy is non-magnetic corresponds to the paramagnetic state (PM), designated by the green color. When magnetic, the value of $q_x$ is depicted by a color between pink and yellow, for $q_x$ from 0.6 to 1 along the $\Gamma-M$ path. This color gradient highlights that we can capture the continuous evolution in \textbf{q} of the order parameter with increasing geometric frustration.

\begin{figure*}
    \includegraphics[width=\linewidth]
    {4-results2D.png}
    \caption{
        Results for magnetic states with wavevectors on the full two-dimensional grid.
        (a)~Phase diagram obtained by extracting the magnetic ordering wavevector that minimizes the total energy $E_\textbf{q}$. The solid colors correspond to the ghost-GA $B=3$ phase diagrams, while the hatched regions represent the phases obtained with GA when they differ from ghost-GA. The green color represents the paramagnetic (PM) state, while the color gradient represents the continuous evolution of magnetic states, with yellow, orange, and red the AFM, spiral quasi-1DAFM states, respectively. As in the results depicted in Fig.~\ref{fig:1D_results}, the magnetic \textbf{q} fall on the high-symmetry path along $\Gamma-X-K-M$. The blue squares and pale blue stars represent boundary anomalies, discussed in the main text and in Fig.~\ref{fig:2d_anomalies}. Certain specific points are discussed in the following subfigures.
        (b-e)~Examples from GA of the total energy $E_\textbf{q}$ and magnetization $M_\textbf{q}$ on the two-dimensional \textbf{q}-grid for representative examples of the (b)~paramagnetic (PM), (c)~quasi-1DAFM, (d)~AFM and (e)~spiral phases. The dashed lines delimit the Brillouin zone. The isolated white points in $E_\textbf{q}$ reveal \textbf{q}-points where the solution did not properly converge.
        }
    \label{fig:2D_vs_Gull}
\end{figure*}

We clearly see that geometric frustration inhibits magnetism and pushes the critical value of $U$ to larger values. In Fig.~\ref{fig:1D_results}(c), we also present three lines for the critical $U$ that separates the magnetic from the non-magnetic ground states: the full line with $\times$ markers, the dashed line with circle markers and the dotted line with open square markers represent GA ($B=1$), ghost-GA ($B=3$) and DMFT extracted from Ref.~\citenum{PhysRevB.94.245145}, respectively. The first two lines are our results, which were obtained by extracting the \textbf{q} ordering wavevectors for $\Delta U=0.1$ around the transition.
Comparing GA and ghost-GA, we find that adding ghost orbitals shifts the phase boundaries towards larger $U$, which is expected as the ghosts allow the variational state to capture some of the dynamical fluctuations~\cite{Giuli2026unifying}.
Therefore, its solution becomes less mean-field-like, reducing the preference towards the ordered phases. 
However, the key result is not just the improvement when including ghosts, but that GA already captures the full topology of the phase diagram. 
Furthermore, comparing $B=3$ ghost-GA with DMFT demonstrates that only a few ghosts suffice for agreement.

\subsection{Two-dimensional phase diagram}
\label{sec:results_2d}

We now explore the entire two-dimensional space of potential magnetic ordering $\textbf{q}$ wavevectors. We do this for three reasons: one is that, given that the triangular lattice supports many nearly degenerate coplanar states, it could potentially have a minimum in $E_\textbf{q}$ anywhere in the full two-dimensional Brillouin zone. The second reason is that we want to compare our results with those obtained with ladder DF in Ref.~\citenum{PhysRevB.107.075106}. 
A third reason is that ghost-GA makes these calculations computationally inexpensive, allowing us to scan the full two-dimensional grid without prohibitive cost.

We use a similar procedure as outlined in the beginning of Sec.~\ref{sec:results_1d}, where we scan the phase diagram by computing the \textbf{q}-dependent total energy $E_\textbf{q}$ and magnetization $M_\textbf{q}$, by employing the rotating spin-frame formulation, but now we explore the whole two-dimensional space of \textbf{q}-vectors. The resulting phase diagram is presented in Fig.~\ref{fig:2D_vs_Gull}(a), for both ghost-GA with $B=3$ (solid colors) and GA (hashed region when disagreeing with ghost-GA). In that figure, the green color depicts the paramagnetic phase (PM). The other colors designate magnetic phases, where the gradient from yellow to red represents the $q_x$ position of the \textbf{q}-vector describing the magnetic order along the $\Gamma-M$ high-symmetry path depicted in Fig.~\ref{fig:1D_results}(a). 
Unsurprisingly, we find that the magnetic phases are easier to stabilize with GA than when including additional ghosts. This behavior is expected, as the additional ghosts capture dynamical fluctuations that are neglected by GA. By going beyond a static-mean-field-like description, they reduce the system’s tendency toward magnetism.

\begin{figure*}
    \centering
    \includegraphics[width=1\linewidth]{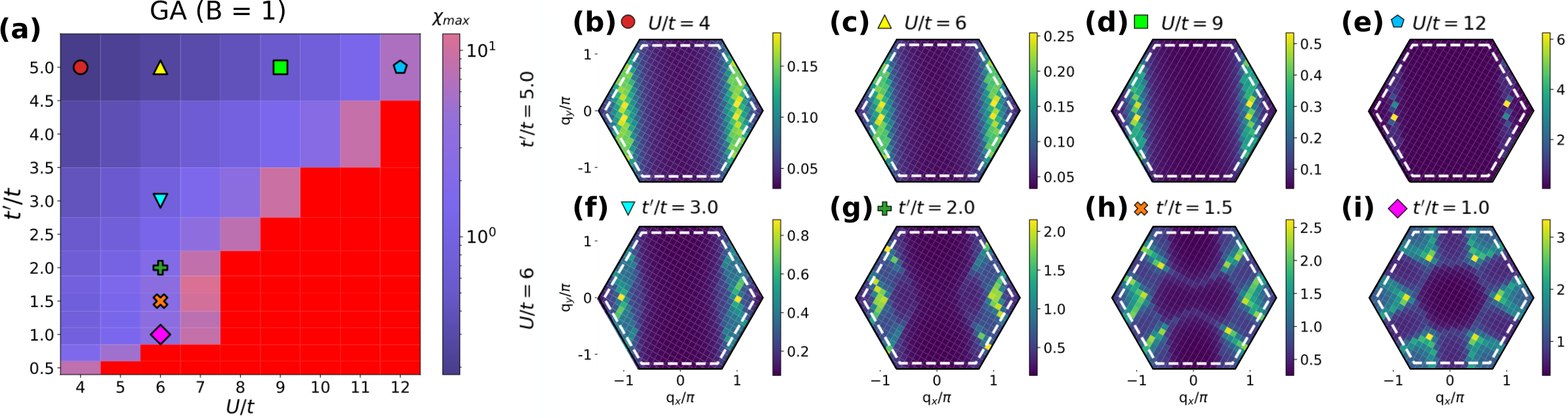}
    \caption{
        Phase diagram based on the magnetic susceptibility.
        (a)~Largest value in \textbf{q} of the susceptibility $\chi^m_\textbf{q}$ defined in Eq.~\ref{eq:magn_susc}, obtain with GA ($B=1$) and using $\phi_x/t=10^{-2}$. As we move away in parameter space from the magnetic dome depicted in Fig.~\ref{fig:2D_vs_Gull}, the amplitude of the susceptibility decreases. In the following subfigures, we present $\chi^m_\textbf{q}$ on the \textbf{q}-grid for a few examples.
        (b-e)~Magnetic susceptibility at fixed $t'/t=5$ along increasing $U/t$. While $\chi^m_\textbf{q}$ starts to look like a vertical line passing through the $X$-point characteristic of the quasi-1DAFM state, the amplitude is far from diverging. As $U/t$ increases, the amplitude also increases, and the maximum of $\chi^m_\textbf{q}$ localizes close to the $X$-point.
        (f-i)~Magnetic susceptibility at fixed $U/t=6$ along decreasing $t'/t$. The maximum of $\chi^m_\textbf{q}$ increases as the geometric frustration is reduced, and the quasi-one-dimensional line of maximum starts folding around the $K$-point.
        }
    \label{fig:2d-susc}
\end{figure*}

\begin{figure}[hb!]
    \centering
    \includegraphics[width=1\linewidth]{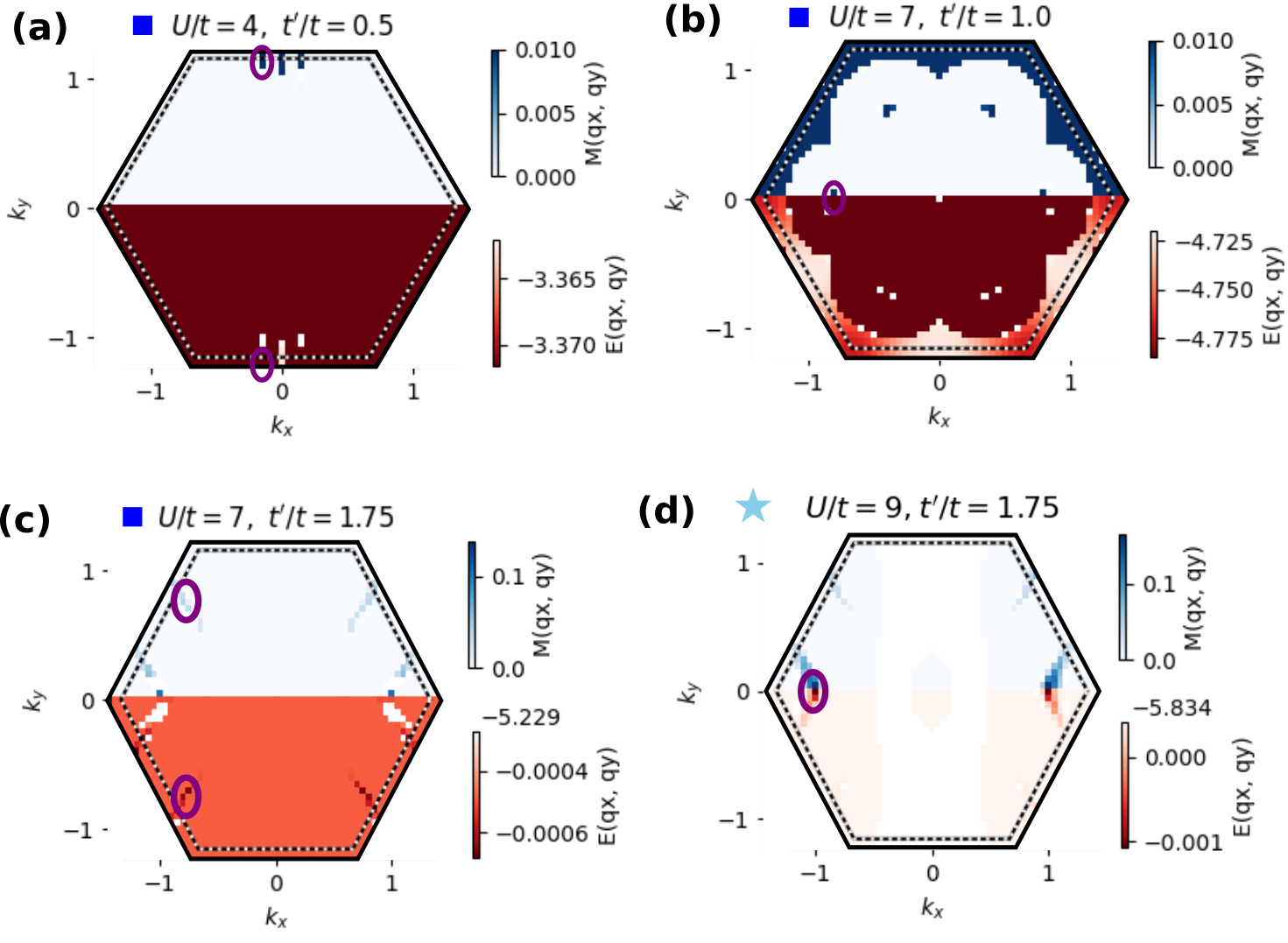}
    \caption{
        Anomalies at the phase boundary between the paramagnetic and magnetic solutions. Some points have a shallow minimum (highlighted by the purple circle) in the \textbf{q}-dependent total energy $E_\textbf{q}$ that is either away from the $\Gamma-M$ high-symmetry path, or at an unexpected place in the $(t'/t,U/t)$ phase diagram. We discard the shallow minima and believe the unexpected points are due to coarse-graining of the \textbf{q}-mesh. Magnetism becomes very sensitive to this at the phase boundary. (a-c) correspond to some points of the GA phase diagram, while (d) corresponds to the ghost-GA phase diagram. }
    \label{fig:2d_anomalies}
\end{figure}

In that figure, a few phases are selected as examples. Their respective total energy $E_\textbf{q}$ and magnetization $M_\textbf{q}$ are presented on the lower and upper halves of the two-dimensional \textbf{q}-grids, respectively, in Figs~\ref{fig:2D_vs_Gull}(b-e).
\begin{itemize}
    \item In (b), $t'/t = 5$ and $U/t = 10$ is an example of a PM state where $E_\textbf{q}$ is independent of \textbf{q} and $M_\textbf{q}$ is vanishing.
    \item In (c), $t'/t = 2.5$ and $U/t = 12$ is an example of a nearly 1DAFM state, with quasi-one-dimensional line of minimum in $E_\textbf{q}$ on the line starting at the $X$-point and perpendicular to the $\Gamma-X$ line. Note however, that the minimum \textbf{q} remains at the $X$-point.
    \item In (d), $t'/t = 0.5$ and $U/t = 8$ is an AFM state, with a pronounced minimum at the $M$-point.
    \item In (e), $t'/t = 1$ and $U/t = 10$ is a spin-spiral state, with a six-fold minimum at the symmetric $K$-point.
\end{itemize}

In addition to the magnetic phases discussed above, we highlight several anomalous points, marked by dark blue squares and pale blue stars in Fig.~\ref{fig:2D_vs_Gull}(a). Selected examples are shown in Fig.~\ref{fig:2d_anomalies}. At the GA phase boundary, the dark blue squares correspond to states with extremely shallow minima in $E_{\mathbf q}$. These minima are not readily resolved by visual inspection and appear as outliers, deviating from the expected $\Gamma-M$ path and instead falling at generic spin-density-wave wavevectors; examples are shown in Fig.~\ref{fig:2d_anomalies}(a-c). We therefore exclude these points from the phase diagram. At the ghost-GA phase boundary, the pale blue stars denote similarly shallow states that produce a non-monotonic phase boundary. These features appear more robust than the dark blue-square anomalies, but may still originate from the coarse graining of the $\mathbf q$ mesh or from the finite number of $\mathbf k$ points used to discretize the non-interacting Hamiltonian. Overall, these anomalies are extremely shallow and occur only near the boundary with the paramagnetic phase. Although they may slightly shift the phase boundary, we do not expect them to affect the qualitative conclusions.

Now, given the phase diagram in Fig.~\ref{fig:2D_vs_Gull}(a), we want to compare it to the one reported in Ref.~\citenum{PhysRevB.107.075106}, which is based on ladder DF calculations. Overall, we argue that the two phase diagrams (DF and ghost-GA) qualitatively agree, apart from the phase labeled 1DAFM reported in their work, which we do not find. We explain the disagreement with two different reasons. 
First, our work considers only single-site ghost-GA, which neglects non-local correlations. In their work, by employing the local vertex to construct the magnetic susceptibility, they diagrammatically consider non-local correlations. To account for those, we would need to perform cluster-ghost-GA calculations, which we plan to investigate in future work.

Second, in our work, we establish the existence of a magnetic state by directly stabilizing it and verifying that its total energy is lower than that of the paramagnetic state. By contrast, Ref.~\citenum{PhysRevB.107.075106} performs the analysis at finite temperature and infers a magnetic phase from the magnetic susceptibility once it exceeds a chosen threshold. Our results suggest that this criterion is not, by itself, sufficient to guarantee that the corresponding phase is stabilized.

In order to examine how far we are from stabilizing the 1DAFM state in the large $t'$, small $U$ region, we
compute the magnetic susceptibility $\chi_{\mathbf q}^{m}$. It is obtained by performing a calculation in the presence of a small field $\phi_x$ along the in-plane $x$ direction, which adds the perturbation 
\begin{equation}
    \delta \hat H_x= \phi_x \sum_i \left(c^\dagger_{i\uparrow}c_{i\downarrow} + c^\dagger_{i\downarrow}c_{i\uparrow}\right),
\end{equation}
resulting in an induced magnetization $M_{\mathbf q}^{\phi_x}$. 
Then, the susceptibility from the normal state is given by
\begin{equation}
    \label{eq:magn_susc}
    \chi^m_\textbf{q} = \lim_{\phi_x \to 0} \frac{M^{\phi_x}_\textbf{q}}{\phi_x}.
\end{equation}

The resulting susceptibility phase diagram obtained with GA ($B=1$) with $\phi_x/t = 10^{-2}$ is presented in Fig.~\ref{fig:2d-susc}(a) through the maximum value of $\chi^m_\textbf{q}$ as a function of \textbf{q}, with a few representative examples in Figs~\ref{fig:2d-susc}(b-i).
In Figs~(b-e), we show how $\chi^m_\textbf{q}$ evolves for fixed $t'/t = 5$, as the interaction $U/t$ is increased. At small $U/t$, the maximum of the susceptibility follows the expected line for the 1DAFM phase, that is, a line that runs vertically and passes through the $X$-point of Fig.~\ref{fig:1D_results}(a). However, the amplitude of that susceptibility is far from diverging. This is understood because the energy scale that is now relevant for the bandwidth is no longer $t$ but instead $t'=5t$, making $U/t'$ very small compared to the bandwidth. As $U/t$ increases, or equivalently here $U/t'$, the magnitude of the susceptibility increases and becomes sharper around the $X$ point, until it is almost divergent.
In Figs~\ref{fig:2d-susc}(f-i), we instead present a scan of $\chi^m_\textbf{q}$ for fixed $U/t=6$, as the geometric frustration $t'/t$ is decreased. At large $t'/t$, $\chi^m_\textbf{q}$ exhibits a line of intensity along the 1DAFM path, and as $t'/t$ decreases, this line starts folding around the $K$ and $K'$-points. Moreover, since $U/t'$ effectively increases, the amplitude of $\chi^m_\textbf{q}$ increases and eventually starts diverging.

Ultimately, the absence of long-range order in the $t/t' \to 0$ limit is not surprising, considering that the Mermin-Wagner theorem prohibits it in a one-dimensional system~\cite{PhysRevLett.17.1133}. However, before reaching this limit, we do not exclude the possibility that such a phase can be stabilized, as suggested in Fig.~\ref{fig:2D_vs_Gull}(c), provided that $U$ is scaled appropriately with $t'$.

\section{Conclusion}
\label{sec:concl}

We have studied the magnetic phase diagram of the anisotropic triangular Hubbard model using GA and its ghost extension within the rotating spin-frame approach, allowing for a direct determination of ordered states via total-energy minimization in momentum space.
A central result of this work is that the standard GA already captures remarkably well the qualitative structure of the phase diagram, including the emergence and evolution of incommensurate magnetic orders across the full range of frustration. Despite its mean-field character, GA correctly identifies the dominant competing phases and their organization in parameter space. This makes it a particularly attractive tool for studying complex magnetic orders in situations where more expensive methods are impractical, such as multi-orbital models, cluster extensions, or systems requiring dense momentum resolution.

At the same time, GA systematically overestimates the stability of ordered phases due to its lack of dynamical fluctuations. We show that introducing ghost orbitals provides a controlled and efficient way to remedy this limitation. Strikingly, only a small number of ghosts ($B = 3$) is sufficient to achieve near-quantitative agreement with DMFT, demonstrating that most dynamical effects can be captured at very low additional cost.
By exploring the full Brillouin zone, we obtain a consistent phase diagram comprising paramagnetic, square antiferromagnetic, spiral, and incommensurate spin-density-wave phases. In contrast to other approaches, we find that the one-dimensional antiferromagnetic phase is not stabilized, even though it appears as a leading instability in certain regimes. 

Looking forward, the framework developed here provides a natural route to investigate magnetic ordering in more realistic and complex settings, including multi-orbital systems, cluster extensions, and materials with spin–orbit coupling, where ordering wavevectors may extend beyond coplanar configurations and involve more intricate spin textures. In this context, GA emerges as a robust and scalable baseline method for exploring incommensurate magnetism, while ghost-GA offers a systematic and computationally efficient pathway toward quantitative accuracy. Together, these approaches enable momentum-resolved studies of complex magnetic states in regimes where fully dynamical methods remain computationally prohibitive.

\begin{acknowledgements}
    The Flatiron Institute is a division of the Simons Foundation.
    N.L. gratefully acknowledges funding from the Simons Foundation (Grant No. 00024037). T.-H.L. gratefully acknowledges funding from the National Science and Technology Council (NSTC) of Taiwan under Grant No. NSTC 112-2112-M-194-007-MY3 and the National Center for Theoretical Sciences (NCTS) in Taiwan. Authors A.K.-M., A.-M.S.T. and M.C. benefit from their RQMP membership https://doi.org/10.69777/309032. A.K-M. and M.C. acknowledge the support of the Natural Sciences and Engineering Research Council of Canada (NSERC), [funding reference number RGPIN-2025-06035]. A.-M.S. T. acknowledges support from the Canada First Research Excellence Fund.
\end{acknowledgements}

\appendix

\section{Mapping between the triangular lattice and the modified square lattice}
\label{app:triangle-to-square}

\begin{figure}[!b]
    \includegraphics[width=.75\linewidth]{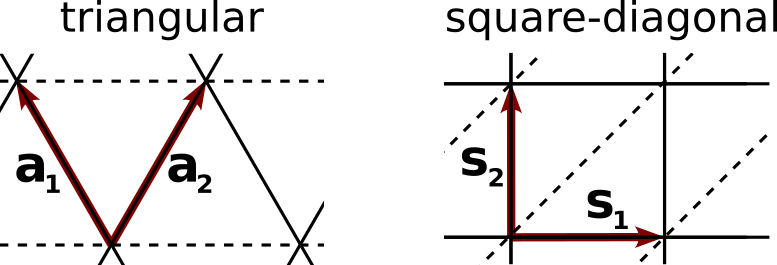}
    \caption{
        Triangular lattice and square lattice with an extra diagonal hopping. Both are topologically equivalent.
    }
    \label{fig:triangle_to_square}
\end{figure}

In this work, we formulate the model on the anisotropic triangular lattice depicted in Fig.~\ref{fig:model}, with primitive vectors shown in Fig.~\ref{fig:triangle_to_square} and given by
\begin{equation}
    \label{eq:triang_latt}
    \mathbf{a}_1^T = \left(-\tfrac{a}{2},\tfrac{a\sqrt{3}}{2} \right), \qquad 
    \mathbf{a}_2^T = \left(\tfrac{a}{2}, \tfrac{a\sqrt{3}}{2}\right).
\end{equation}
Nearest-neighbor hopping \(t\) acts along \(\pm \mathbf{a}_1\) and \(\pm \mathbf{a}_2\), while the third bond \(\pm (\mathbf{a}_1 - \mathbf{a}_2)\) carries the hopping \(t'\). This is the standard anisotropic triangular lattice, and all our results are expressed in this representation. To compare directly with Ref.~\cite{PhysRevB.94.245145}, we also make use of the equivalent square-diagonal description, also shown in Fig.~\ref{fig:triangle_to_square}, defined by the primitive vectors
\begin{equation}
    \mathbf{s}_1^T = (a,0), \qquad 
    \mathbf{s}_2^T = (0,a),
\end{equation}
with hopping \(t\) along $\pm \mathbf{s}_1$ and $\pm \mathbf{s}_2$, and hopping \(t'\) along \(\pm(\mathbf{s}_1+\mathbf{s}_2)\).

A linear change of basis relates the two representations. We define a linear map \(\mathbf{L}\) acting on coordinates in the square lattice such that
\begin{equation}
\mathbf{L}\,\mathbf{s}_1 = \mathbf{a}_1, \qquad
\mathbf{L}\,\mathbf{s}_2 = \mathbf{a}_2.
\end{equation}
This map is represented by the matrix
\begin{equation}
\mathbf{L}
=
\frac{1}{2}\left(
\begin{array}{cc}
-1 & 1 \\
\sqrt{3} & \sqrt{3}
\end{array}
\right),
\end{equation}
so that for any position of a site \(\mathbf{R}_\square = m\,\mathbf{s}_1 + n\,\mathbf{s}_2\) on the square-diagonal lattice, we obtain the corresponding position on the triangular lattice as
\begin{equation}
\mathbf{R}_\triangle = \mathbf{L}\,\mathbf{R}_\square.
\end{equation}
The determinant of \(\mathbf{L}\) is \(\det \mathbf{L} = -\tfrac{\sqrt{3}}{2} \neq 0\), so the map is invertible and provides a one-to-one correspondence between the two lattices. Under this mapping, the three bond directions $\mathbf{s_1}$, $\mathbf{s}_2$ and $\mathbf{s}_1 + \mathbf{s}_2$ of the square-diagonal lattice are sent to the three bond directions \(\mathbf{a}_1\), \(\mathbf{a}_2\) and \(\mathbf{a}_1 - \mathbf{a}_2\) of the triangular lattice, preserving the connectivity and thus the Hamiltonian up to this linear distortion.

In the rotating spin-frame, magnetic order is parameterized by a wavevector \(\mathbf{q}\). On the square-diagonal lattice, the commonly used cut \(q_x = q_y\) imposes equal phase change along two \(t\)-bonds, i.e.
\begin{equation}
\mathbf{q} \cdot \mathbf{s}_1 = \mathbf{q} \cdot \mathbf{s}_2.
\end{equation}
To express the same physical condition on the triangular lattice, we require equal phase change along \(-\mathbf{a}_1 \) and \(\mathbf{a}_2\) (two $t$-bonds with a $t'$ bond in between):
\begin{equation}
-\mathbf{q} \cdot \mathbf{a}_1  = \mathbf{q} \cdot \mathbf{a}_2.
\end{equation}
Writing \(\mathbf{q} = (q_x, q_y)\) in Cartesian coordinates and inserting
Eq.~(\ref{eq:triang_latt}), we obtain
\begin{equation}
\tfrac{a}{2}q_x - \tfrac{\sqrt{3}a}{2}q_y = \tfrac{a}{2}q_x + \tfrac{\sqrt3a}{2}q_y
\quad\Longrightarrow\quad
q_y = 0.
\end{equation}
Thus, the square-lattice line \(q_x = q_y\) is exactly equivalent to the \(\Gamma \rightarrow M\) path
\begin{equation}
q_y = 0
\end{equation}
in the Brillouin zone of the triangular lattice (shown in Fig.~\ref{fig:1D_results}(a)), and all Brillouin-zone cuts used in this work are presented in this triangular-lattice parametrization.

\end{document}